\documentclass[aps,prb,floatfix,showpacs,twocolumn,groupedaddress,notitlepage,noshowpacs]{revtex4}

\usepackage{graphics}
\usepackage{graphicx}
\usepackage{epsfig}
\usepackage{amssymb}
\usepackage{amsmath}
\usepackage{amstext}
\usepackage{bm}
\usepackage{times}
\usepackage{textcomp}

\usepackage{epstopdf}

\renewcommand{\sp}{{sp}}

\renewcommand{\vec}[1]{\mathbf{#1}}

\newcommand{\ER}{\ensuremath{E_\text{rec}}}
\newcommand{\intvecr}{\hspace{-0.05cm} \int \hspace{-0.7ex} d^3r \hspace{0.05cm} }

\newcommand{\ket}[1]{\ensuremath{|#1\rangle}}

\newcommand{\uu}{{\uparrow\uparrow}}
\newcommand{\ud}{{\uparrow\downarrow}}

\newcommand{\AFFILP}{\affiliation{Institut f\"ur Laser-Physik, Universit\"at Hamburg,
Luruper Chaussee 149, 22761 Hamburg, Germany} }

\begin{document}

\title{Quantum phase transition to unconventional multi-orbital superfluidity in optical lattices}

\author{Parvis Soltan-Panahi}
\altaffiliation{Both authors contributed equally to this work.}
\author{Dirk-S\"oren L\"uhmann}
\altaffiliation{Both authors contributed equally to this work.}
\author{Julian Struck}
\author{Patrick Windpassinger}
\author{Klaus Sengstock}
\AFFILP

\begin{abstract}
Orbital physics plays a significant role for a vast number of important phenomena in complex condensed matter systems such as high-T$_c$ superconductivity and unconventional magnetism. In contrast, phenomena in superfluids -- especially in ultracold quantum gases -- are commonly well described by the lowest orbital and a real order parameter\cite{Pitaevskii2003}. Here, we report on the observation of a novel multi-orbital superfluid phase with a \textit{{complex}} order parameter in binary spin mixtures. In this unconventional superfluid, the local phase angle of the complex order parameter is continuously twisted between neighboring lattice sites. The nature of this twisted superfluid quantum phase is an interaction-induced admixture of the p-orbital favored by the graphene-like band structure of the hexagonal optical lattice used in the experiment. We observe a second-order quantum phase transition between the normal superfluid (NSF) and the twisted superfluid phase (TSF) which is accompanied by a symmetry breaking in momentum space. The experimental results are consistent with calculated phase diagrams and reveal fundamentally new aspects of orbital superfluidity in quantum gas mixtures. Our studies might bridge the gap between conventional superfluidity and complex phenomena of orbital physics. 
\end{abstract}

\maketitle

\begin{figure}
\includegraphics[width=2\linewidth]{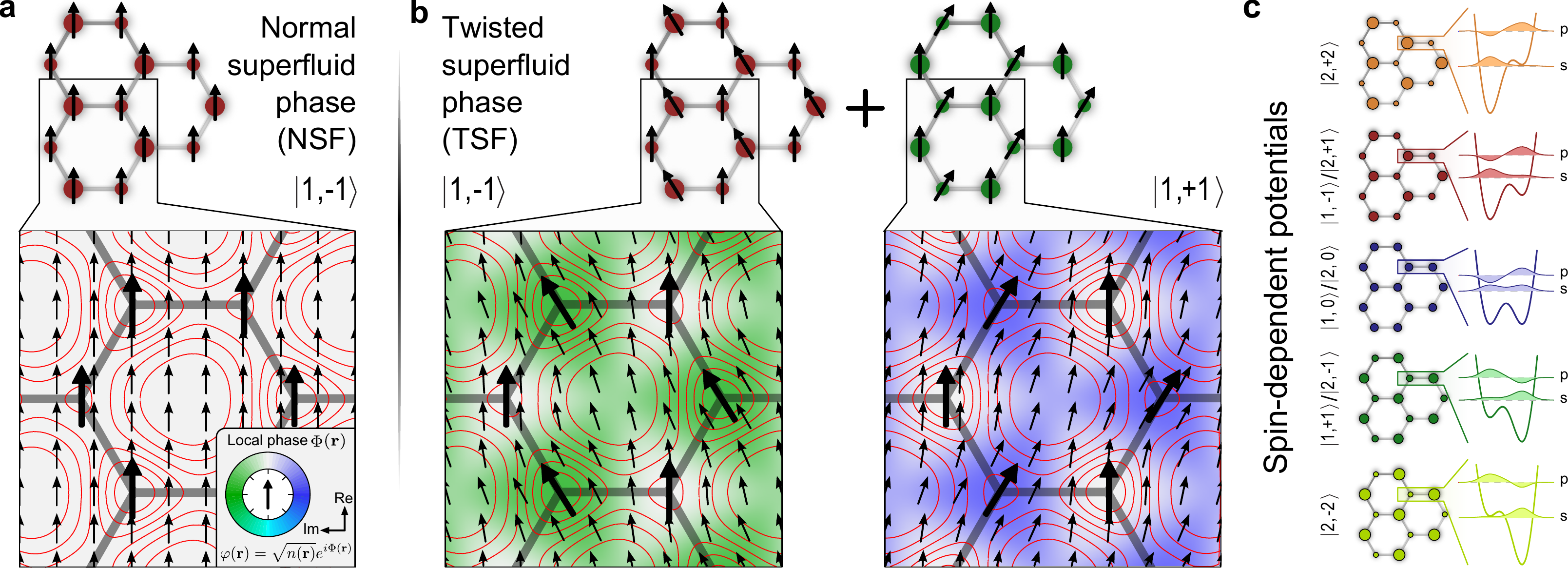}
\begin{minipage}[t]{2\linewidth}
 \caption{Normal and twisted superfluid phase. \rm
The contour plots depict the superfluid order parameter $\varphi(\mathbf{r})=\sqrt{n(\mathbf{r})} e^{i\Phi(\mathbf{r})}$ (red lines) in the hexagonal lattice. The arrows and color coding represent the local phase angles $\Phi(\mathbf{r})$. \textbf{a}, Normal superfluid: Single-component spin systems exhibit a real superfluid order parameter with constant local phases. \textbf{b}, Twisted superfluid: In binary spin mixtures, the local phases $\Phi(\mathbf{r})$ vary continuously between the sites of the hexagonal lattice leading to a complex superfluid order parameter for both components. This twisted phase is caused by a complex admixture of the p-band. \textbf{c}, The spin-dependent lattice causes a density modulation in dependence on the spin state $\ket{F,m_F}$. The s- and p-orbital wave functions as well as the spin-dependent potentials are shown as a one-dimensional cut along two adjacent lattice sites.  
}
 \end{minipage}
\end{figure}

The topological properties of graphene and its remarkable band structure have recently opened a new field in physics\cite{Geim2007}. The linear dispersion relation at the Dirac points proves to be a fascinating key aspect of this material as it gives rise to phenomena such as quasi-relativistic particles\cite{Du2009} and an anomalous quantum Hall effect\cite{Novoselov2005,Zhang2005}. The possibility to realize hexagonal optical lattices\cite{SoltanPanahi2011} permits the emulation of graphene-like physics with ultracold atoms\cite{ Zhu2007,Wu2008,Lee2009}. In particular, loading bosonic particles in such lattices renders completely new possibilities such as studying next-nearest neighbor processes and tunneling blockades in multi-component systems\cite{SoltanPanahi2011}. In general, optical lattices have proven to be a versatile tool to simulate Hubbard-like systems and actively drive and monitor quantum phase transitions \cite{Greiner2002, Jordens2008,Schneider2008,Struck2011,Bloch2008}. The important role of higher orbitals has recently been demonstrated for quantum gas mixtures in the strongly correlated regime \cite{Will2010,Best2009,Luhmann2008,Lutchyn2009}. However, for weakly interacting systems such as superfluids, higher orbitals are generally expected to have only marginal effects. So far, orbital superfluidity has been observed only in {\it excited} states with limited lifetimes\cite{Wirth2011,Olschlager2011,Muller2007}.

\begin{figure*}
\includegraphics[width=1\linewidth]{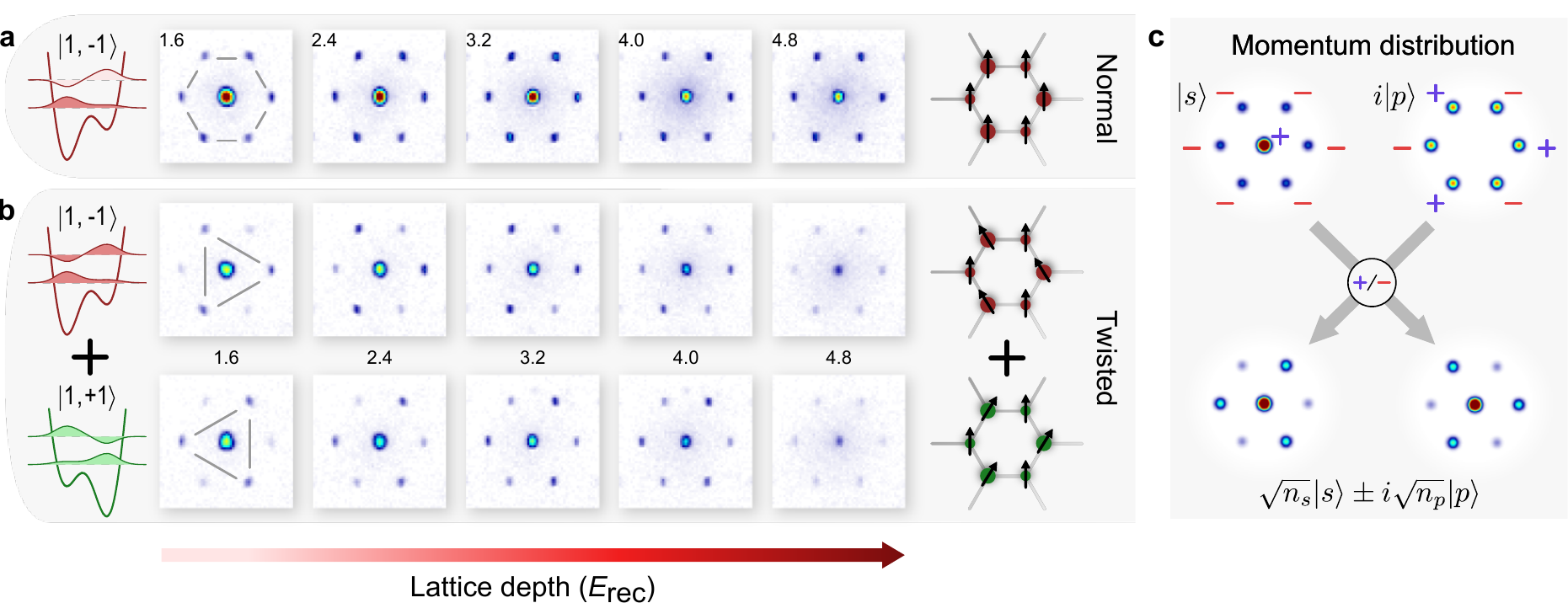}
\caption{Symmetry breaking in momentum space. \rm
\textbf{a}, The normal superfluid phase shows a six-fold-rotational symmetry due to the lattice geometry. Averaged TOF images are shown for different lattice depths (indicated in the images). \textbf{b}, For certain binary mixtures, an interaction-induced reduced three-fold rotational symmetry is clearly observed for both constituents being separated by a Stern-Gerlach field. The triangular pattern is opposite for the two constituents. The six-fold rotational symmetry is restored for large lattice depths. \textbf{c}, Calculated momentum distributions for s- and p-orbital wave functions for $\ket{F,m_F=0}$, where the p-orbital wave function in momentum space has an alternating sign in its first-order peaks. A complex superposition of both orbitals leads to the observed TOF pattern for the TSF phase even for small admixtures of the p-orbital (shown for p-band admixtures of $n_p=1-n_s=0.05$).
}
\end{figure*}

Here, we demonstrate the realization of a bosonic superfluid {\it ground-state} where higher orbital physics plays a crucial role. In conventional superfluids (Fig.~1a), the local phase angle $\Phi(\mathbf{r})$ of the order parameter 
\begin{equation}
	\varphi(\mathbf{r})=\sqrt{n(\mathbf{r})} e^{i\Phi(\mathbf{r})}
\end{equation}
is constant (represented by arrows) and $\varphi(\mathbf{r})$ can therefore be chosen as real. In contrast, the observed TSF ground-state reveals a non-trivial complex superfluid order parameter, where the phase factor $e^{i\Phi(\mathbf{r})}$ is continuously twisted in the complex 

\vspace{10.164cm}
~

\noindent  plane (Fig.~1b). As we will demonstrate, this unconventional behavior arises from a strong interaction-induced coupling of  s- and p-orbitals. Most strikingly, even a quantum phase transition between the NSF and the TSF phase is directly observed in our experiment. It is driven by the competition between intra- and interspecies interactions in binary mixtures. For its realization, we use a mixture of ultracold atoms in two spin states $\ket{F,m_F}$ of the hyperfine manifold $F=1,2$ with Zeeman states $m_F$ (in our experiment $^{87}$Rb atoms\cite{Schmaljohann2004}, see methods). The repulsively interacting atoms are confined in a spin-dependent hexagonal optical lattice \cite{SoltanPanahi2011}. 

As a central aspect, the formation of the twisted superfluid phase originates from both the spin-dependency and the specific topology of the hexagonal lattice. The topology leads to a graphene-like band-structure with the particular feature that s- and p-bands are separated only on the order of the tunneling energy. In addition, the spin-dependency induces an individual sublattice structure for different $\ket{F,m_F}$ states. This leads to an alternating density modulation for $\ket{F,m_F\neq 0}$ spin states (see Fig.~1c), which substantially alters the interspecies interactions. The combination of both effects causes a strong coupling of s- and p-bands for the case of spin mixtures. 

In the following, we first explain how the TSF phase can be identified experimentally and subsequently discuss theoretical phase diagrams as well as experimental results in detail. The NSF phase possesses the expected six-fold rotational symmetry in momentum space. This is observed in experiments via time-of-flight (TOF) imaging as exemplarily shown in Fig.~2a for the single-component spin-state $\ket{1,-1}$. In stark contrast, the twisted superfluid phase is accompanied by a symmetry breaking in momentum space which appears as an alternating pattern in the first order momentum peaks (Fig.~2b). This reduced three-fold rotational symmetry reflects the occurrence of a twisted complex phase factor $e^{i\Phi(\mathbf{r})}$, which we observe only for mixtures of two spin states. Figure 2b shows the results for a $1\!:\!1$ mixture of $\ket{1,-1}$ and $\ket{1,+1}$ atoms, where the spin states are separated in the experiment by a Stern-Gerlach field. The occurrence of the twisted superfluid phase is clearly visible for very low lattice depths. Here, both components show a complementary momentum distribution reflecting the opposite phase twist as indicated in Fig.~1b. For increasing lattice depths, the transition to the normal superfluid phase is observed by the restoration of the six-fold rotational symmetry (Fig.~2b). Finally, the overall interference contrast vanishes as the system approaches the Mott insulator phase, where the atoms localize on individual lattice sites.

\begin{figure*}
\includegraphics[width=1\linewidth]{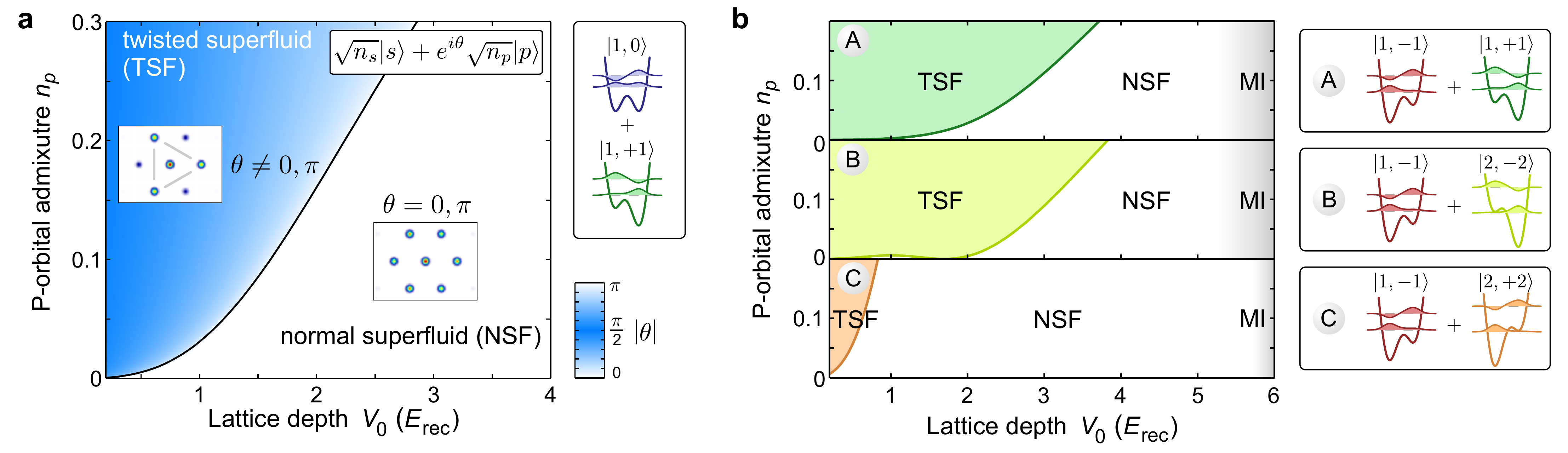}
\caption{Quantum phase transition to a twisted superfluid phase. \rm
 \textbf{a}, Zero-temperature phase diagram for the $\ket{1,0}$ state in admixture with the $\ket{1,+1}$ state. The diagram is parameterized by the relative occupation of the p-orbital $n_p=1-n_s$ and the hexagonal lattice depth $V_0$. 
The white area defines the normal superfluid phase (NSF) and the blue area the twisted superfluid phase (TSF), where the color encoding represents the value of the order parameter $\theta$. \textbf{b}, Phase diagram of $\ket{1,-1}$ in admixture with (A) $\ket{1,+1}$, (B) $\ket{2,-2}$ and (C) $\ket{2,+2}$.  For a lattice depth of $V_0=5$-$6 \ER$, the system undergoes the transition to a Mott insulator (MI), where the entrance points depend on the particular spin mixture\cite{SoltanPanahi2011}.
}
\end{figure*} 

Remarkably, the clear signature of the TSF phase persists even for p-band admixtures as small as a few percent, which is a typical value in our experiment. This relies on the fact that the first order momentum peaks of the p-orbital are much stronger than those of the s-orbital leading to a strong amplification of the p-band contributions, which can be clearly identified in the experimental signal (see Fig.~2c). The observed alternating pattern for the twisted superfluid is caused by the quantum interference of s- and p-band contributions, which permits an extremely sensitive probing of the local phase properties via time-of-flight. 

The connection between the hybridization of s- and p-orbitals and the transition to the TSF phase can be explained as follows. In general, a superposition between s-orbital $\ket{s}$ and p-orbital $\ket{p}$ can be written as
\begin{equation}
\ket{\varphi} = \sqrt{n_s} \ket{s} + e^{i\theta} \sqrt{n_p} \ket{p},
\end{equation}
where the coefficients $n_s$ and $n_p$ denote the fraction of atoms in the s- and the p-orbital, respectively. The global phase angle $\theta$ between both orbitals is crucial for the formation of the twisted superfluid. It takes the value which minimizes the energy of the system and can lead to two different physical situations: For $\theta=0$ (or $\pi$), the system is in the normal superfluid phase, where no interference takes place and the alternating momentum pattern vanishes. In contrast, $\theta\neq0$ causes a destructive interference of the first order peaks (Fig.~2c) thereby revealing the twist of the local phases $\Phi(\mathbf{r})$. Thus, the global phase angle $\theta$ takes the important role of an order parameter describing the phase transition between NSF and TSF, where the TSF phase is defined by a non-zero value of $\theta$. 

We explore this phase transition theoretically using a multi-band mean-field approach (see methods), which leads to the phase diagrams presented in Fig.~3. The phase diagrams show the results for different binary spin mixtures. In agreement with the experimental results, the twisted superfluid emerges only in shallow lattices. It is important to mention that in our hexagonal optical lattice configuration, different spin states preferably occupy different sublattices and therefore the interplay of intra- and interspecies interaction strongly depends on the spin mixture considered. In particular, Fig.~3b demonstrates that the TSF phase area is drastically reduced for spin mixtures predominately occupying the same sublattice (mixture C) in comparison to spin mixtures, where each component occupies a different sublattice (mixture A and B). In addition, the occurrence of the TSF phase depends on the admixture of the p-band orbital. This also explains the absence of the TSF phase for single-component samples (Fig.~2a) where the population of the p-orbital is negligible.
 
In the following, we experimentally investigate the phase diagrams above. As an experimental indicator characterizing the NSF and TSF phase, we define a {\it triangular interference contrast} $I_\triangle$ which is illustrated in Fig.~4. It serves as a measure of the order parameter $\theta$ of the NSF-TSF transition, where $I_\triangle \neq 0$ corresponds to the TSF phase. For spin mixture A, the TSF phase is clearly resolved for $V_0<4\ER$ (Fig.~4a). As expected for symmetry reasons, both components exhibit the same triangular contrast $|I_\triangle|$. In accordance with the phase diagrams presented in Fig.~3, mixture B exhibits a similar behavior as mixture A, where in both cases different sublattices are occupied by the constituents (Fig.~4b). The substantial difference of the predicted TSF phase areas for mixtures B and C is also clearly revealed in our experiment.

To gain further insight into the underlying processes of the NSF-TSF transition, we turn back to its theoretical description. The observed quantum phase transition is entirely driven by the competition between intra- and interspecies interactions and can thus occur at zero temperatures. We apply mean-field theory, where we restrict our analysis to an effective two-mode Hamiltonian\cite{Zhou2010} (see methods). For the superfluid order parameter, we consider s- and p-band contributions described by Eq.~(2), where the order parameter of the transition $\theta$ is the relative phase angle between the two orbitals. In the calculation, higher bands and non-zero quasi-momentum states can be neglected to first order.

\begin{figure}
\includegraphics[width=1\linewidth]{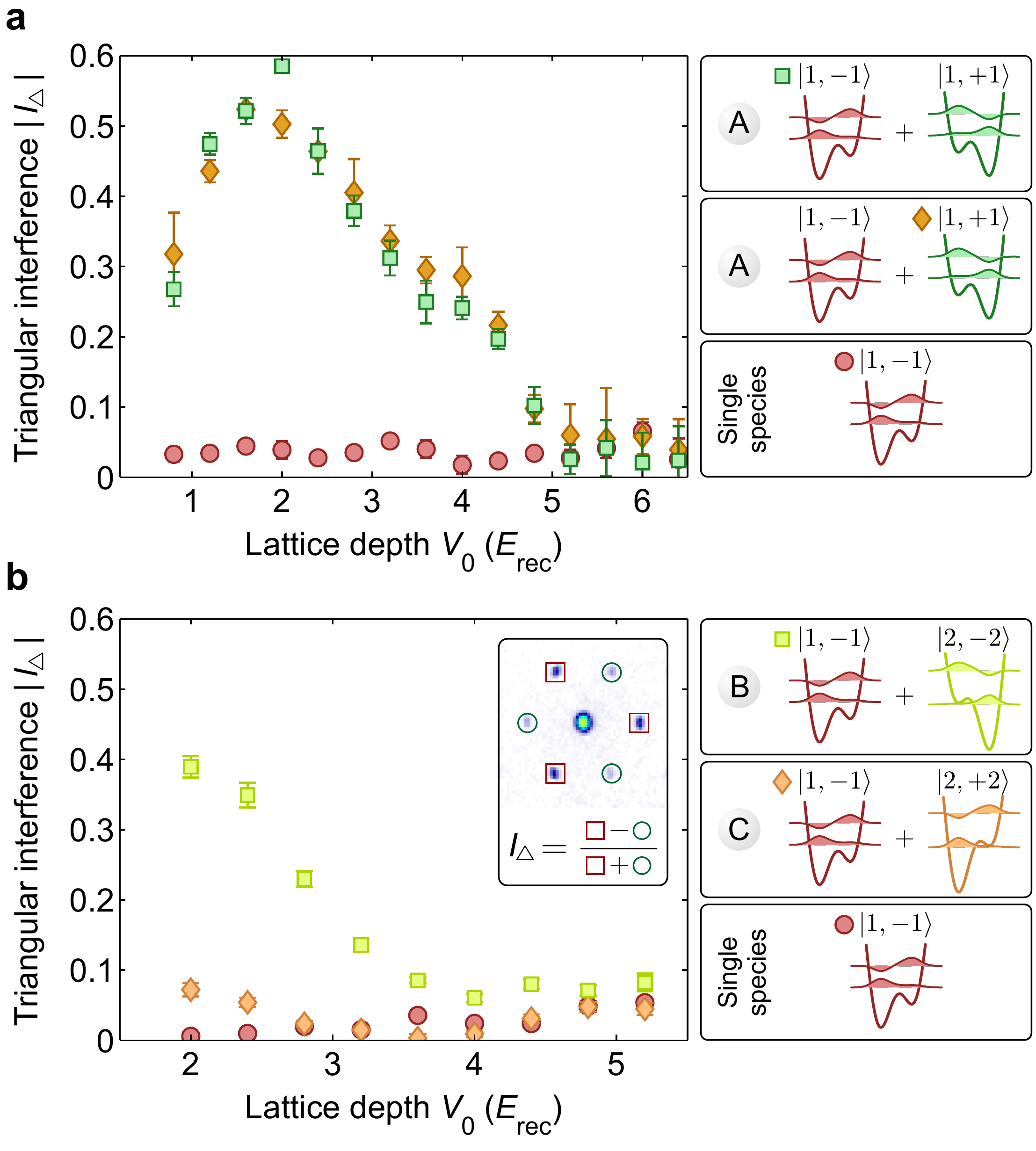}
\caption{Observation of the NSF-TSF quantum phase transition. \rm
Absolute value of the triangular interference contrast $I_\triangle$ (see inset) as a function of the lattice depth $V_0$ for \textbf{a}, spin mixture A and \textbf{b}, spin mixtures B and C. The TSF phase is identified by a non-zero value of $I_\triangle $, whereas $I_\triangle$ vanishes in the NSF phase. The error bars indicate the standard deviation of the mean.
}
\end{figure}

For simplicity, we discuss here a mixture of $\ket{1,0}$ with $\ket{1,-1}$ atoms (see Fig.~3a). In this case, the symmetry of s- and p-band wave functions leads to only two competing $\theta$-dependent terms in the energy functional, namely, the interspecies interaction $v_{sp}(\theta)=2N^2V_\sp\,\sqrt{n_s n_p} \cos(\theta)$ and the intraspecies interaction $w_\sp(\theta)=2N^2W_\sp\, n_s n_p \cos(2\theta)$. Here, $N$ is the number of particles in each component, $V_\sp$ and $W_\sp$ are integrals of s- and p-orbital wave functions with $V_\sp<0$ and $W_\sp>0$ (see methods). For sufficiently small admixtures of the p-band $n_p$, the interspecies interaction $v_\sp$ dominates and therefore $\theta=0$ minimizes the energy, which corresponds to a normal superfluid. However, for a critical value of $n_p$ the energy functional $v_\sp(\theta)+w_\sp(\theta)$ no longer exhibits a minimum at $\theta=0$. 
As a central result, this defines the phase boundary $n_p^\text{crit}$ of the NSF-TSF transition, which is given for the considered case by
\begin{equation}
n_p^\text{crit}=\frac{1}{2} - \sqrt{\frac{1}{4} - \left( \frac{ V_{sp}}{4 W_{sp}} \right)^2}. 
\end{equation}
The phase transition is of second order as the second derivative of the energy functional is discontinuous at this boundary. In the applied theory, the Hamiltonian is invariant under the transformation $\theta\to -\theta$ causing a two-fold degenerate ground-state in the twisted phase. However, in the experiment always the same of the two ground-states is observed. Further studies are necessary to investigate this non-spontaneous symmetry breaking. 

Our study of a new type of complex superfluid phase paves the way for further investigations of the interplay between orbital physics and strong correlations. In particular, a possible competition between the twisted superfluid and the strongly correlated Mott insulator phase can be realized by increasing the interactions, e.g. by means of Feshbach resonances. Moreover, further insight into the roles of intra- and interspecies interactions can be gained using binary mixtures consisting of two different atomic states, where both interactions differ considerably from each other. In addition, dynamically driven phase transitions may be observable in our systems, by preparing a dynamical superposition of s- and p-orbitals for one of the two spin components by microwave coupling.

\subsection*{Methods}
\small

\textit{Creation of spin-dependent hexagonal lattices --}
The spin-dependent hexagonal lattice is realized by intersection of three coplanar laser beams under an angle of 120 degrees. The laser beams are derived from a Ti:Sapphire laser operated at $830\text{nm}$ (red detuned), where each beam is linearly polarized within the plane of intersection. Orthogonal to the plane, we apply a retro-reflected one-dimensional lattice at $V_{\text{1D}}=8.8\ER$ operated at the same wavelength (for details see Ref.~\onlinecite{SoltanPanahi2011}) For the experimental parameters, we expect an admixture of the p-orbital between zero (for vanishing interactions) and a few percent for large lattice depths. 

\textit{Preparation and detection schemes for spin mixtures --}
We start with a Bose-Einstein condensate of typically several $10^5$ atoms in the stretched state $\ket{1,-1}$, which is confined in a nearly isotropic crossed dipole trap ($\omega \approx 2\pi \times 90 \mathrm{Hz}$). The preparation of the different pure and mixed spin states is performed with aid of radio-frequency and/or microwave sweeps. After the state preparation we ramp up the optical lattice within $80\text{ms}$ using an exponential ramp. Within the ramping time the coherence between different spin states is lost.
To separate different spin-components during $27\text{ms}$ time-of-flight, a Stern-Gerlach gradient field is applied prior to absorption imaging.

\textit{Theoretical model --}
To first order, one spin species experiences the interaction with the non-interacting density $M|\varphi'(\mathbf{r})|^2$
of the other spin species with $M$ atoms. Thus, we can write the effective Hamiltonian for one spin state as
\begin{equation}
  \hat H \!\!=\!\!\intvecr \hat\psi^\dagger(\mathbf r) \!\left[
       H_0 + g_\ud M|\varphi'(\mathbf{r})|^2 
     + \frac{g_\uu}{2} \hat\psi^\dagger(\mathbf r) \hat\psi(\mathbf r)
  \right]\! \hat\psi(\mathbf r)
\label{H}
\end{equation}
where $ H_0=\frac{{\mathbf p}^2}{2m}+ V(\vec r)$ is the operator for kinetic and potential energy and $\hat\psi$ is the bosonic field operator. The intra- and interspecies interaction parameters are labeled as $g_\sigma=\frac{4\pi \hbar^2}{m} a_\sigma$ with $\sigma=\uu,\ud$, respectively, and $a_\sigma\approx100a_0$. For shallow lattices, we assume that only s- and p-band Bloch functions $\varphi_{s,p}$ with quasi-momentum $q=0$ contribute. For large particle numbers and weak interactions, we apply mean-field theory and expand the field operators according to Eq.~(2)
\begin{equation}
\hat\psi(\mathbf r) \to \sqrt{n_s} \varphi_s + e^{i\theta} \sqrt{n_p} \varphi_p,
\end{equation}
where $\varphi_{s,p}$ are real functions.  
The energy functional can be divided in a $\theta$-independent and a  $\theta$-dependent part, where the latter is given by $H_\theta(n_p,\theta)= v_\sp + w_\sp + x_\sp$ with
\begin{equation}
\begin{split}
v_\sp&=2MNV_\sp\, \sqrt{n_s n_p} \cos(\theta),\\ 
w_\sp&=2N^2W_\sp\, n_s n_p \cos(2\theta),\\
x_\sp&=4N^2 \left( X_s n_s + X_p n_p \right) \sqrt{n_s n_p} \cos(\theta).
\end{split}
\end{equation}
These terms depend on the interspecies integral $V_{sp} = g_{\ud} \intvecr |\varphi'|^2\, \varphi_s^* \varphi_p$, the intraspecies integrals $W_{sp} = \frac{g_\uu}{2} \intvecr \varphi_s^{*2} \varphi_p^2$, $X_s = \frac{g_\uu}{2} \intvecr \varphi_s^* |\varphi_s|^2 \varphi_p$ and $X_p = \frac{g_\uu}{2} \intvecr \varphi_s^* |\varphi_p|^2 \varphi_p$. The latter two vanish for $m_F=0$ due to reasons of parity. The phase boundary of the phase transition between normal and twisted superfluid phase is defined by 
\begin{equation}
 	2W_{sp} \sqrt{n_p (1-n_p)} = \left| \frac{M}{2N} V_{sp} + X_s (1-n_p) + X_p n_p \right|.
\end{equation}
When approaching the Mott insulator transition, higher quasi-momentum states become occupied and the two-mode description presented here is no longer fully valid. This could explain quantitative deviations between theory and experiment.   
The momentum distribution $\rho_\text{TOF}$ can be calculated using the Fourier transformed wave functions $\tilde\varphi_{s,p}$ (see Fig.~2c)
\begin{equation}
	\rho_\text{TOF}=|\tilde\varphi_s|^2 n_s+|\tilde\varphi_p|^2 n_p 
	      + 2\sqrt{n_s n_p}\,\text{Re}\left(\tilde\varphi_s^* \tilde\varphi_p e^{i\theta}  \right).
\end{equation}
While for $\theta=0$ or $\pi$ the third terms vanishes, it causes an interference effect for other phase angles $\theta$.

\subsection*{Acknowledgments}
\small
The work has been funded by DFG grants FOR 801 and GRK 1355 as well as by the Landesexzellenzinitiative Hamburg which is supported by the Joachim Herz Stiftung.

\end{document}